\documentclass[aps,prb,two column,superscriptaddress]{revtex4-1}

\usepackage{gensymb}
\usepackage{hyperref}
\usepackage{graphicx}
\usepackage{amsmath}
\usepackage{amssymb}
\usepackage{bm}
\usepackage{color}
\usepackage{bookmark}
\usepackage{tabularx}
\usepackage{mathtools}
\usepackage{microtype}
\usepackage{cancel}
\usepackage{relsize}
\usepackage{textcomp}
\usepackage{ulem}

\begin{document}
\newcommand{\Ss}{\textsubscript}
\newcommand{\Us}{\textsuperscript}
\newcommand{\STO}{SrTiO\Ss{3}}
\newcommand{\GTO}{GdTiO\Ss{3}}
\newcommand{\TO}{TiO\Ss{2}}
\newcommand{\ELxTO}{Eu\Ss{1-x}La\Ss{x}TiO\Ss{3}}
\newcommand{\ELTO}{Eu\Ss{0.9}La\Ss{0.1}TiO\Ss{3}}
\newcommand{\GTOx}{GdTiO\Ss{3+x}}
\newcommand{\LAO}{LaAlO\Ss{3}}
\newcommand{\LGAO}{La\Ss{1-x}Gd\Ss{x}AlO\Ss{3}}
\newcommand{\GLTO}{Gd\Ss{1-y}La\Ss{y}Ti\Ss{z}Al\Ss{1-z}O\Ss{3}}
\newcommand{\LTO}{LaTiO\Ss{3}}
\newcommand{\RTO}{RETiO\Ss{3}}
\newcommand{\NGO}{NdGaO\Ss{3}}
\newcommand{\LCO}{LaCrO\Ss{3}}
\newcommand{\PyroCl}{Re\Ss{2}Ti\Ss{2}O\Ss{7}}
\newcommand{\ETO}{EuTiO\Ss{3}}
\newcommand{\ELaTO}{Eu\Ss{0.9}La\Ss{0.1}TiO\Ss{3}}
\newcommand{\ETOPyro}{Eu\Ss{2}Ti\Ss{2}O\Ss{7}}
\newcommand{\ox}{O\Ss{2}}
\newcommand{\tit}{Ti\Us{3+}}
\newcommand{\tif}{Ti\Us{4+}}
\newcommand{\DC}{\degree C}
\newcommand{\mx}{\times}
\newcommand{\RAN}{$R_{xy}^{AHE}$}
\newcommand{\dxy}{3d\Ss{xy}}
\newcommand{\dxyz}{3d\Ss{xz/yz}}
\newcommand{\dxyzo}{d\Ss{xz/yz}}
\newcommand{\dxyo}{d\Ss{xy}}
\newcommand{\rec}{~\cite}
\newcommand{\Rahe}{$R_{AHE}$}

\title{On the resistance minimum in \LAO /\ELxTO /\STO{} heterostructures}


\author{N. Lebedev}
 \email{lebedev@physics.leidenuniv.nl}
\affiliation{Huygens-Kamerlingh Onnes Laboratory, Leiden University, P.O. Box 9504, 2300 RA Leiden, The Netherlands}
\author{Y. Huang}
\affiliation{Van der Waals-Zeeman Institute, University of Amsterdam, Science Park 904, 1098 XH Amsterdam, The Netherlands}
\author{A. Rana}
\affiliation{Center for Advanced Materials and Devices, BML Munjal University (Hero Group), Gurgaon, India - 122413}
\affiliation{MESA+ Institute for Nanotechnology, University of Twente, P.O. Box 217, 7500 AE Enschede, The Netherlands}
\author{D. Jannis}
\author{N. Gauquelin}
\author{J. Verbeeck}
\affiliation{Electron Microscopy for Materials Science, University of Antwerp, Campus Groenenborger Groenenborgerlaan 171, 2020 Antwerpen, Belgium}
\author{J. Aarts}
 \email{aarts@physics.leidenuniv.nl}
\affiliation{Huygens-Kamerlingh Onnes Laboratory, Leiden University, P.O. Box 9504, 2300 RA Leiden, The Netherlands}

\date{\today}

\begin{abstract}
In this paper we study \LAO{}/\ELxTO{}/\STO{} structures with  nominally x = 0, 0.1 and different thicknesses of the \ELxTO{} layer. We observe that both systems have many properties similar to previously studied \LAO/\ETO/\STO{} and other oxide interfaces, such as the formation of a 2D electron liquid for 1 or 2 unit cells of \ELxTO{}; a metal-insulator transition driven by the thickness increase of \ELxTO{} layer; the presence of an Anomalous Hall effect (AHE) when driving the systems above the Lifshitz point with a backgate voltage; and a minimum in the temperature dependence of the sheet resistance below the Lifshitz point in the one-band regime, which becomes more pronounced with increasing gate voltage. However, and notwithstanding the likely presence of magnetism in the system, we do not attribute that minimum to the Kondo effect, but rather to the properties of \STO{} crystal and the inevitable effects of charge trapping when using back gates.
\end{abstract}


\maketitle

\section{Introduction\label{intro}}

At oxide interfaces, mainly based on the \STO{} (STO), two-dimensional electron liquids (2DELs) are easily formed. These systems have significant freedom of manipulating various magnetotransport properties by applying a gate voltage. One outstanding feature, of interest for spintronics applications, is the tunability of the spin polarisation or magnetism. Its signature, the Anomalous Hall effect (AHE), was extensively studied in the past decade in various STO-based structures\rec{JoshuaPNAS,NMATLAOETO, GunkelPRX, ZhangLSMOPRB, GanPRB, ChrisNPHYS, ZhangACS, GTOpaper, ParkNCOM}. It was shown, in particular, that the AHE can be completely switched off by a gate voltage, at least in some of this structures. The switching point is strongly connected to the so-called Lifshitz transition\rec{JoshuaNCOM}, which separates two regimes with only the \dxy{} band occupied, leading to single band transport, or with involvement of both the \dxy{} and \dxyz{} bands. The latter regime coincides with the presence of AHE and, therefore with the occurrence of spin-polarisation\rec{JoshuaPNAS,JoshuaNCOM,NMATLAOETO,ZhangLSMOPRB,GunkelPRX,GanPRB}. They are absent in the one band regime but there a non-monotonous temperature dependence of the sheet resistance, with a pronounced minimum has been observed. This behavior has been attributed to the presence of a Kondo-regime\rec{NMATLAOETO, ZhangLSMOPRB}.
For STO(001)-based interfaces, it has been proposed that different coupling between the localized magnetic moments and the \dxy{} and \dxyz{} electrons, due to the different orbital orientation relatively to the interface plane, can explain various magnetotransport properties of STO-structures, including a gate tunable Kondo-like minimum, AHE, and behavior of the in-plane magnetoresistance (MR)\rec{JoshuaPNAS, RuhmanPRB}. Nevertheless, the mechanism for invoking Kondo-like effects is still a matter of debate. It was argued that enhanced spin-orbit coupling (SOC) at low temperature can produce some features of the in-plane MR which previously have been interpreted as supporting Kondo behaviour\rec{DiezPRL}.

This work aims to study the gate-tunable magnetic interactions. We choose as a starting point the well known delta-doped system \LAO /\ETO /STO (LAO/ETO/STO). This system, where the 2DEL presumably forms at the interface between the ETO and the LAO, is reported to be ferromagnetic\rec{LucaPRB, NMATLAOETO} and to exhibit both tunable AHE and Kondo-like behavior. Bulk \ETO{} is an antiferromagnetic band insulator\rec{ChienPRB,McGuireJAP,SchiemerPRB,BHPRB,BessasPRB, EllisPRB,KatsufujiPRB}. The oxidation state of Eu is Eu\Us{2+}; therefore, Ti is in the tetravalent state and does not contribute to the magnetization. However, doping the bulk with La will lead to the development of ferromagnetism\rec{KatsufujiPRBELTO, ELTOPRL}, because La is in the La\Us{3+} state, and the doping introduces electrons into the conduction band of the $3d$ $t_{2g}$ states of the Ti, turning ETO in ferromagnetic metal\rec{KatsufujiPRBELTO, ELTOPRL}. In our work, to understand the role of La doping (or diffusion) on the magnetic interactions, we also investigated nominally non-magnetic LAO/STO and LAO/\ELTO{} /STO (LAO/ELTO/STO) We find a gate-tunable AHE by inserting the 2 unit cells (u.c.) of ETO or ELTO, and we find a transition between tunable AHE and Kondo-like regimes. However, unlike the AHE, the Kondo-like regime seems to be present even at zero gate voltage, although the dip becomes more pronounced at finite voltages.
\begin{figure}[b]
\includegraphics[scale=0.7]{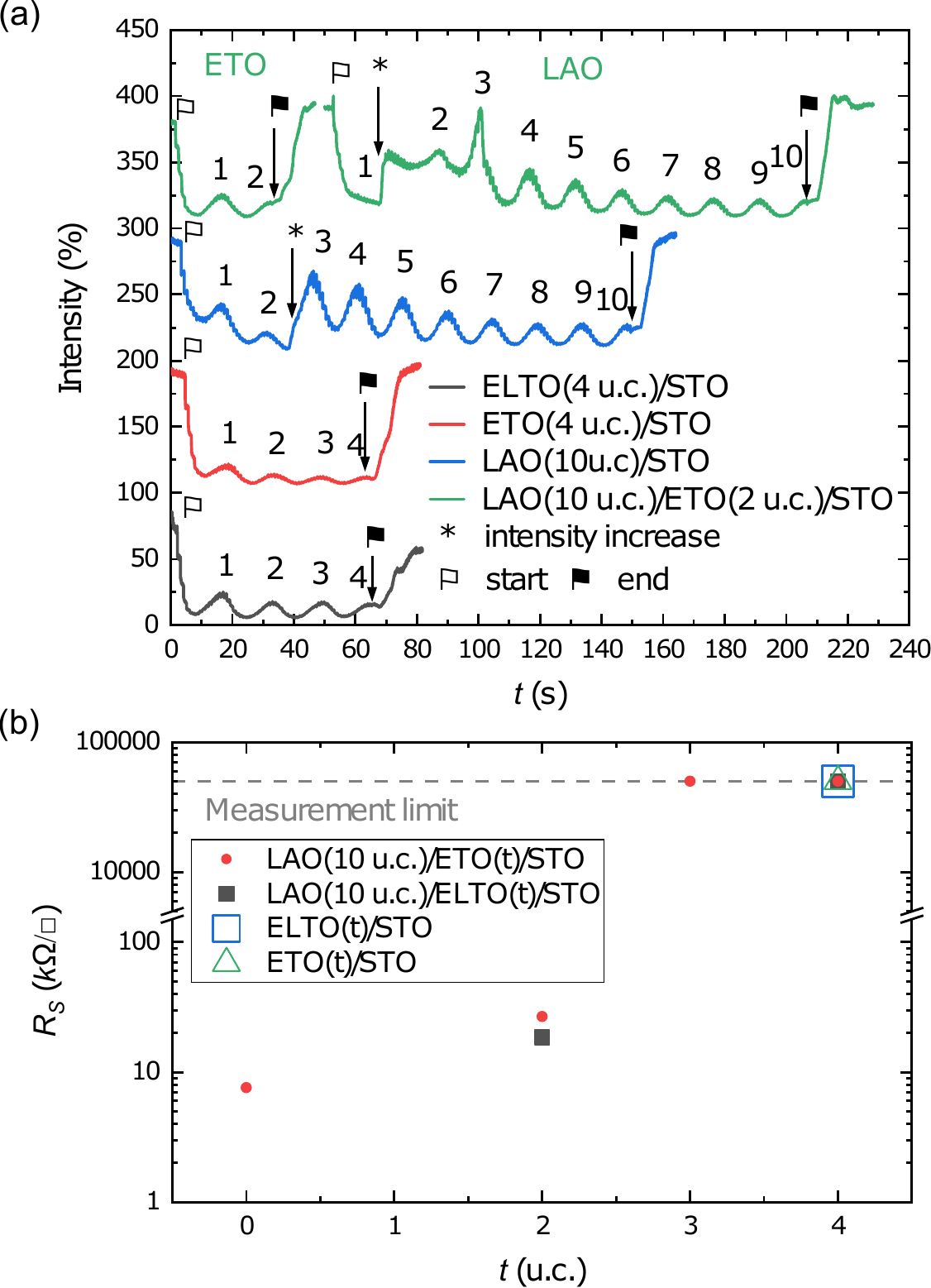}
\caption{(a) RHEED intensity monitoring during the growth of (from bottom to top) ELTO(4)/STO, ETO(4)/STO, LAO(10)/STO and LAO(10)/ETO(2)/STO. (b) The thickness dependence of the sheet resistance $R_S$ at 300 K for E(L)TO and LAO/E(L)TO samples grown on STO. Note the metal-to-insulator transition above 2 unit cells. }\label{Fig1-ETO_MIT}
\end{figure}
\begin{figure*}[!]
\includegraphics[scale=0.47]{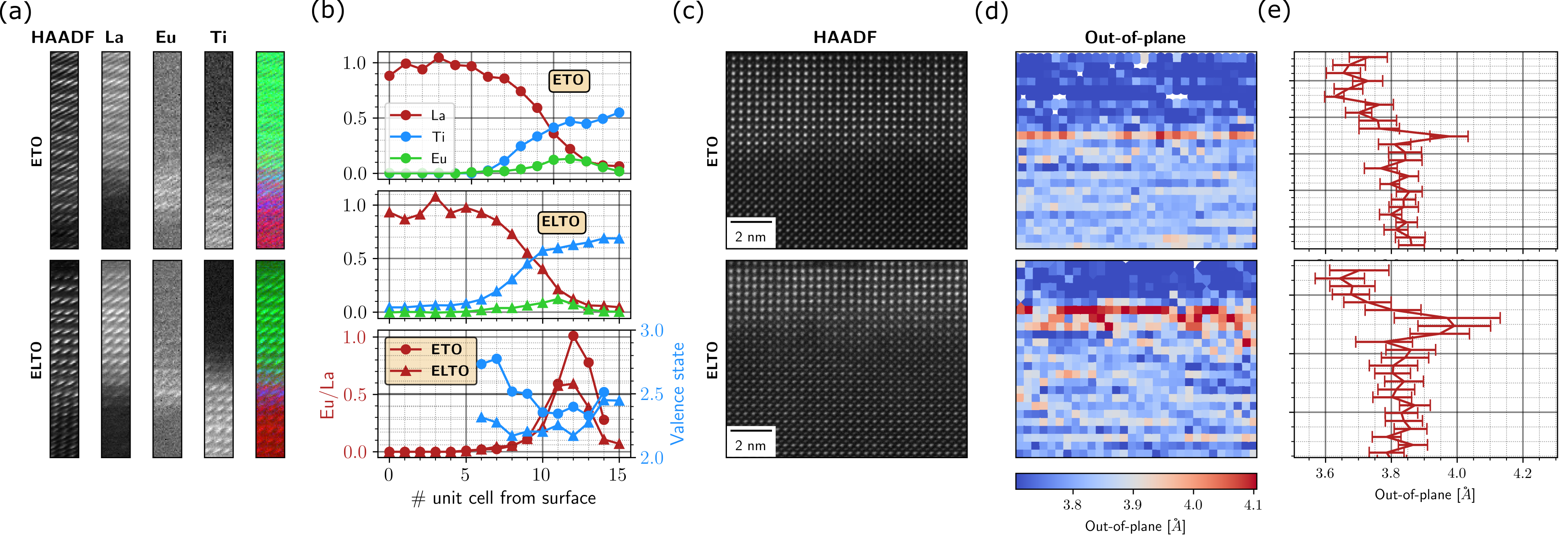}
\caption{(a) High-Angle Annular Dark-Field imaging of the ETO (upper panels) and the ELTO (lower panels) samples. The presence of La, Eu, Ti when crossing the interface is indicated. The color plot show La in green, Eu in purple, Ti in red. (b) EELS analysis of the La, Eu, Ti content as function of unit cell distance from the surface for ETO (upper panel) and ELTO (middle panel). The lower panel shows the Eu/La ratio (left-hand scale) and the Eu valence state (right-hand scale) for both samples. (c)  HAADF imaging of the interface region for ETO and ELTO. (d) and (e) Analysis of the out-of- plane lattice constant from the image in (c). LAO/ETO/STO is denominated as ETO and LAO/ELTO/STO as ELTO.}\label{Fig2-ETO_TEM}
\end{figure*}
We come to the conclusion that the voltage dependent Kondo-like resistance behavior is not a sign of magnetic interactions, but most likely the result of the interplay between the electron trapping mechanism and the temperature dependence of the STO permittivity, the carrier concentration, and various scattering mechanisms. The combination of these effects leads to more effective back gating at low temperatures and therefore to an increase in resistance which has little to do with magnetic scattering.

\section{Experimental Details\label{EXPER}}

The oxide structures were grown by Pulsed Laser Deposition (PLD) on \TO-terminated $(001)$-oriented STO. In the literature, two types of targets have been used to grow ETO, the pyrochlore material \ETOPyro \rec{LucaPRB, TanakaJMR, ChaeJOE, HatabayashiJJAP, FujitaAPL, ShimamotoAPL}, and the perovskite material \ETO \rec{KUGIMIYAJMMM, JiangPCCP, WangHHAPL,RZhaoAPL}. In this work, we choose to work with the latter. The PLD targets for ETO and \ELaTO{} were fabricated from sintered and pressed powders. The LAO target was commercially purchased. The nominal La doping was chosen at $10\% $ to ensure a significant difference between intentionally doped films and unintentional doping due to possible intermixing. However, as we will show in Section \ref{TEM} the resulting doping was different. Significant optimization of the growth parameters was required to obtain good growth. In order to prevent bulk conductivity of the STO due to oxygen depletion at the high growth temperatures, some oxygen in the PLD chamber is needed, but the oxygen pressure should not be too high; otherwise the ETO and ELTO films become amorphous or form the pyrochlore structure.

Based on the optimization of the growth parameters we chose a fluence of $1.54$ J/cm\Us{2} for ETO and ELTO and spot size $1.38$ mm\Us{2}. For LAO a growth fluence of $1.3$ J/cm\Us{2} and a spot size of $1.76$ mm\Us{2} were used. The temperature was set at $800$ \DC{} and the nominal pressure, consisting of a 1:1 mix of Ar:O\Ss{2}, was set at $1 \mx 10^{-4}$ mbar. The maximum thickness of \ELTO{} layer was fixed at 4 u.c.'s. Above that value, growth of the \ELTO{} films result mostly in an amorphous RHEED pattern; still, that thickness is enough to study delta-doped LAO/STO structures. The following samples were grown (the numeral denoting number of u.c.'s): LAO(10)/ETO(t)/STO (with t= 2, 3, 4), LAO(10)/ELTO(t)/STO (with t=2, 4), ELTO(4)/STO, ETO(4)/STO and LAO(10)/STO.

Figure~\ref{Fig1-ETO_MIT}a shows the RHEED intensity variations during growth for both ETO and ELTO, which are quite similar. Oscillations are clearly visible for the ETO(4)/STO and ELTO(4)/STO samples (black and red lines). The RHEED pattern for ETO(4)/STO shows additional lines at the 1/2 position between the main lines for films with $t> 2$ u.c. (insert in Supplement Fig.~S1b); such lines have been observed for stoichiometric \ELTO{} films\rec{ShkabkoAPLM}. The STO surface steps can been clearly seen in the Atomic Force Microscopy (AFM) scan before and after growth (Supplement Fig.~S1), but the edges become more rough than on the bare STO surface. Moreover, there are island-like features in the topography of the film. During the growth of LAO, clear oscillations are seen when growing directly on STO (blue line in Fig.~\ref{Fig1-ETO_MIT}a). The RHEED pattern shows a transition from 2D growth to 3D growth (insert in Supplement Fig. S1c), most likely due to the low oxygen pressure. A clear underlying STO step pattern can be seen in the AFM scan in Supplement Fig.~S1c. This is different when the LAO layer is grown on ETO or ELTO films. In Fig.~\ref{Fig1-ETO_MIT}a (green line) the first two oscillations during LAO growth become less pronounced, and their intensity decreases with the increase of the ETO or ELTO layer thickness. The thickness of the LAO layer was determined from the number of pulses for a single period of oscillations and, strictly speaking, can vary between $8$ and $10$ u.c. That variation is not crucial for the conductivity of LAO/\ETO /STO, in contrast to the thickness of the ETO layer\rec{LucaPRB}. The LAO layer grown on the ETO or ELTO also shows 3D features in the RHEED pattern, but the 2D features are more pronounced (insert in Supplement Fig. S1d). The underlying topography of ETO/ELTO films can still been seen after depositing the LAO layers as shown in Supplement Fig. S1d. After growth, the magnetotransport properties were measured in van der Pauw geometry in a PPMS from Quantum Design with home-built insert for performing gating experiments. Two samples of LAO(10)/ETO(2)/STO and LAO(10)/ELTO(2)/STO were subsequently analysed by Scanning Transmission Electron Microscopy (STEM),  Electron Energy Loss Spectroscopy (EELS) and High-Angle Annular Dark-Field (HAADF) imaging. \\
The basic behavior of our samples is similar to the ones studied in Refs. \onlinecite{LucaPRB,NMATLAOETO}. In particular, both our LAO/ETO/STO and LAO/ELTO/STO show a Metal-Insulator transition (MIT) as function of E(L)TO thickness around 2 u.c. (Figure~\ref{Fig1-ETO_MIT}b), which is in agreement with the results of Ref. \onlinecite{LucaPRB}. From now, we will refer to the two conducting samples, LAO(10)/\ETO(2)/STO and LAO(10)/\ELaTO(2)/STO, as LAO/ETO/STO and LAO/ELTO/STO respectively. The fact that uncapped ETO and ELTO samples were insulating above 2 u.c.'s indicates that La doping of the E(L)TO layer in our films is not the driving mechanism of the MIT \rec{KatsufujiPRBELTO, ELTOPRL}. At room temperature, LAO/STO has a lower sheet resistance $R_S$ than LAO/ELTO/STO, which in turn is lower than that of LAO/ETO/STO.
\begin{figure*}
\includegraphics[scale=0.19]{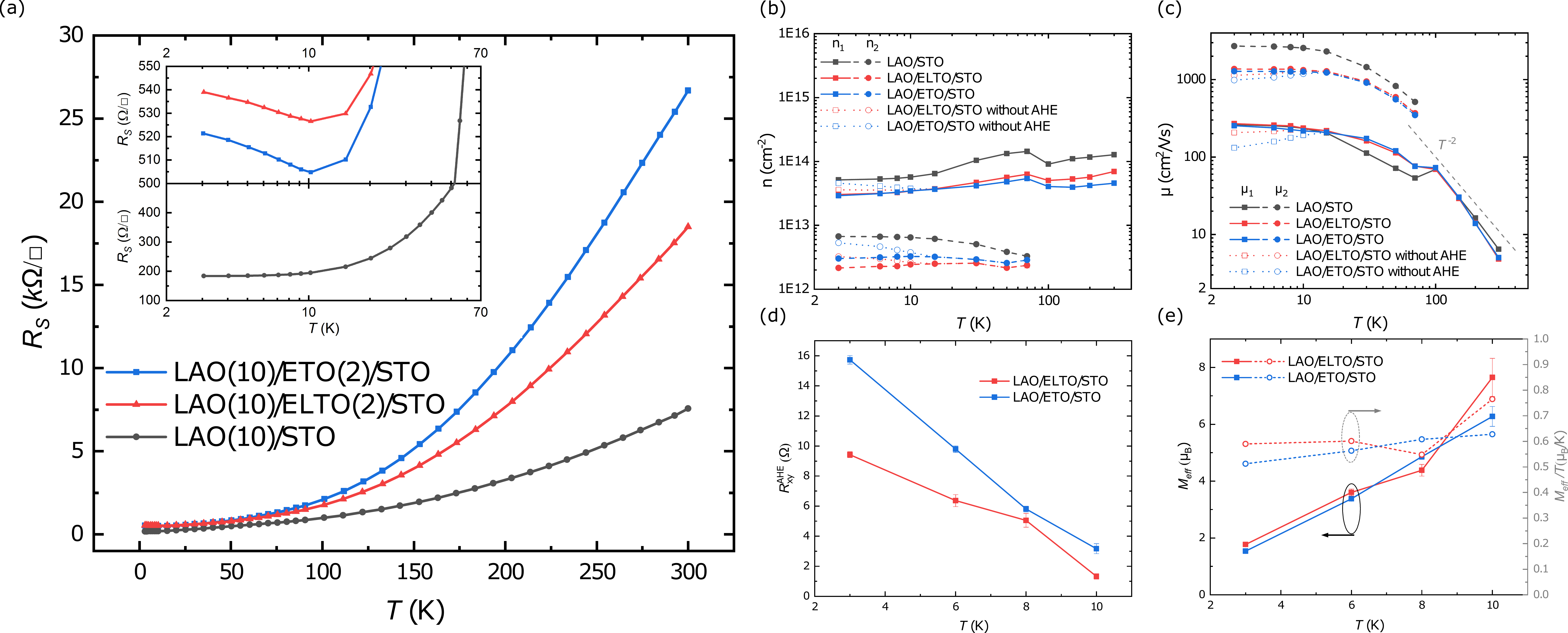}
\caption{(a) Temperature dependence of $R_S$ for samples with and without an E(L)TO interlayer. The insert shows details of the low-temperature region. For the LAO/STO and the LAO/E(L)TO/STO samples, temperature dependences are given for (b) the carrier concentrations $n_i$, (c) the mobilities $\mu_i$, (d) the Anomalous Hall coefficient, and (e) values of $M_{eff}$ (filled symbols) plus the ratio $M_{eff}/T$ ratio (open symbols). They were taken from fitting as described in the Supplementary. Note that $n_i$ and $\mu_i$ are given both with and without correction for AHE when appropriate. \label{Fig3-ETO_RS}}
\end{figure*}

\section{Results\label{Results}}
\subsection{TEM characterisation\label{TEM}}

Extensive HAADF and EELS analysis (Fig. \ref{Fig2-ETO_TEM}a-b) revealed clear intermixing of La and Eu in the ETO sample. The sample shows a higher Eu/La ratio,  which means less La, than the ELTO sample. This is to be expected, but the fact remains that the difference between both samples is smaller than was intended. For both samples, the effective ELTO layer becomes 3 unit cell thick due to diffusion and the Ti diffuses 5 unit cells inside the LAO. We also note a higher diffusion of Eu inside the LAO for ETO than ELTO. This is evident from the presence of a higher concentration gradient in LAO/ETO/STO. Some Eu diffuses inside the LAO, where it is present as Eu\Us{3+}, which is nonmagnetic. Overall, the Eu valence states in the ETO and ELTO layers are Eu\Us{2.36+} and Eu\Us{2.3+} respectively (see the blue line in the bottom panel of Fig.\ref{Fig2-ETO_TEM}b and also Supplement Fig. S2). Thus, the doping layers appear to resemble each other quite closely. However, Eu is distributed rather uniformly in the ELTO layer, which is not the case in the ETO layer where Eu is concentrated close to the interface. That may lead to a higher concentration of magnetic moments in the first layer next to the interface of ETO and STO. HAADF image (Fig. \ref{Fig2-ETO_TEM}c) was used to extract lattice parameter variation for both samples, which are present in Fig. \ref{Fig2-ETO_TEM}d-e. In both cases, the films are relaxed in-plane, as shown in Supplement Fig. S2c-f. We can notice an expansion of the out-of-plane lattice parameter within one unit cell for ETO while it is present over 3 unit cells in the ELTO sample.

\subsection{Magnetotransport properties at zero gate voltage}
\label{ETO_MT}

Before applying gate voltages, we analyze the temperature dependence of the magnetotransport properties for the three conducting samples.
The introduction of the thin sheet of \ELTO{} also changes the temperature dependence of $R_S$: the LAO/STO sample exhibits a monotonous decrease down to 3 K, whereas the delta-doped samples show an upturn below 10~K (Fig.~\ref{Fig3-ETO_RS}a). Such a change in $R_S(T)$ has been reported in Ref.\onlinecite{LucaPRB} and was attributed to a possible Kondo effect\rec{BrinkmanNMAT, LucaPRB}. The behaviour of the MR is described in the Supplementary.

The behaviour of the Hall coefficient is of more significance in the context of this work. Below 100~K, the Hall coefficient $R_H$ develops a non-linearity in high fields, indicating the onset of two-band transport\rec{JoshuaNCOM} (see Supplement Fig.~S3). Furthermore, below 10 K, there is also a low-field non-linearity in the magnetically doped samples. This indicates the presence of an Anomalous Hall Effect (AHE), which has often been observed and analysed\rec{JoshuaNCOM, NMATLAOETO, GunkelPRX, GTOpaper} in order to determine the carrier concentrations $n_i$ and mobilities $\mu_i$of both bands, and the Anomalous Hall coefficient \Rahe{}. Our analysis uses a substraction method described earlier in Ref.~\rec{GTOpaper}. Details are included in the Supplementary, where we also introduce an effective magnetic moment $M_{eff}$ to take account of the presence of the AHE.\\
\begin{figure*}
\includegraphics[scale=0.22]{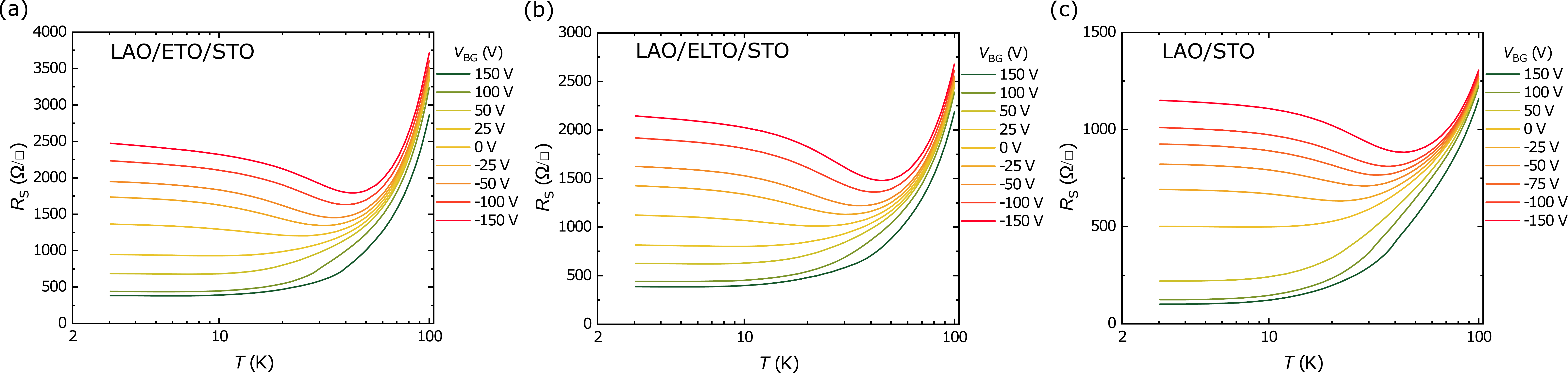}
\caption{Temperature dependence of the sheet resistance $R_S$ for different applied gate voltages in a range between +150~V and -150~V, as indicated, for (a) LAO/ETO/STO, (b) LAO/ELTO/STO, and (c) LAO/STO.}\label{Fig4-ETO_gating} 
\end{figure*}
The extracted values for $n_i, \mu_i$, \Rahe{}, and  $M_{eff}$ are shown in Fig.~\ref{Fig3-ETO_RS}b-e. Also shown are the values for $n_i, \mu_i$ when no AHE is taken into account. We see that in the one band region, the mobility of the samples roughly follows a $T^{-2}$ dependence in the temperature range 150-50 K (Fig.~\ref{Fig3-ETO_RS}c). The temperature dependence above 150 K is often ascribed to longitudinal optical phonon scattering, and the origin of $T^{-2}$ mobility dependence is usually attributed to electron-electron scattering\rec{TrierJPD,MikheevAPL}. As was mentioned above, the second band appears below 100 K. Some unphysical jumps are probably related to the low carrier concentration of the second band leading to overestimating (underestimating) the carrier concentration (mobility) for both type of carriers. Overall it is clear that the trend is an increase of $\mu_{1,2}$ towards low temperatures with saturation below 10 K and a continuous although small decrease of $n_1$. The saturation of the mobility at a low-temperature limit is most likely due to the interface, to ionized donors, and to ionized impurity scattering\rec{TrierJPD}. In the fit without the AHE, with decreasing temperature we see an increase in carrier concentration and an decrease in mobility for the magnetically doped samples below 10 K. However, when accounting for the AHE, the fits become consistent with the general trend of LAO/STO. The AHE grows when lowering the temperature (Fig. \ref{Fig3-ETO_RS}d), but we see that the fit parameter $M_{eff}$ decreases almost linearly with temperature as can be seen better from the ratio $M_{eff}/$T (Fig. \ref{Fig3-ETO_RS}e). This is quite counterintuitive. If $M_{eff}$ would simply represent the saturation magnetization of a ferromagnetically ordering interface, the trend should be opposite. Since an ordering temperature of around 10~K is expected, this points to a different interpretation of $M_{eff}$, which may not be surprising, since the AHE depends on more than the magnetization alone, with SOC to start with. It does not affect the determination of \Rahe{} and the shape of the function used.
\begin{figure*}
\includegraphics[scale=0.24]{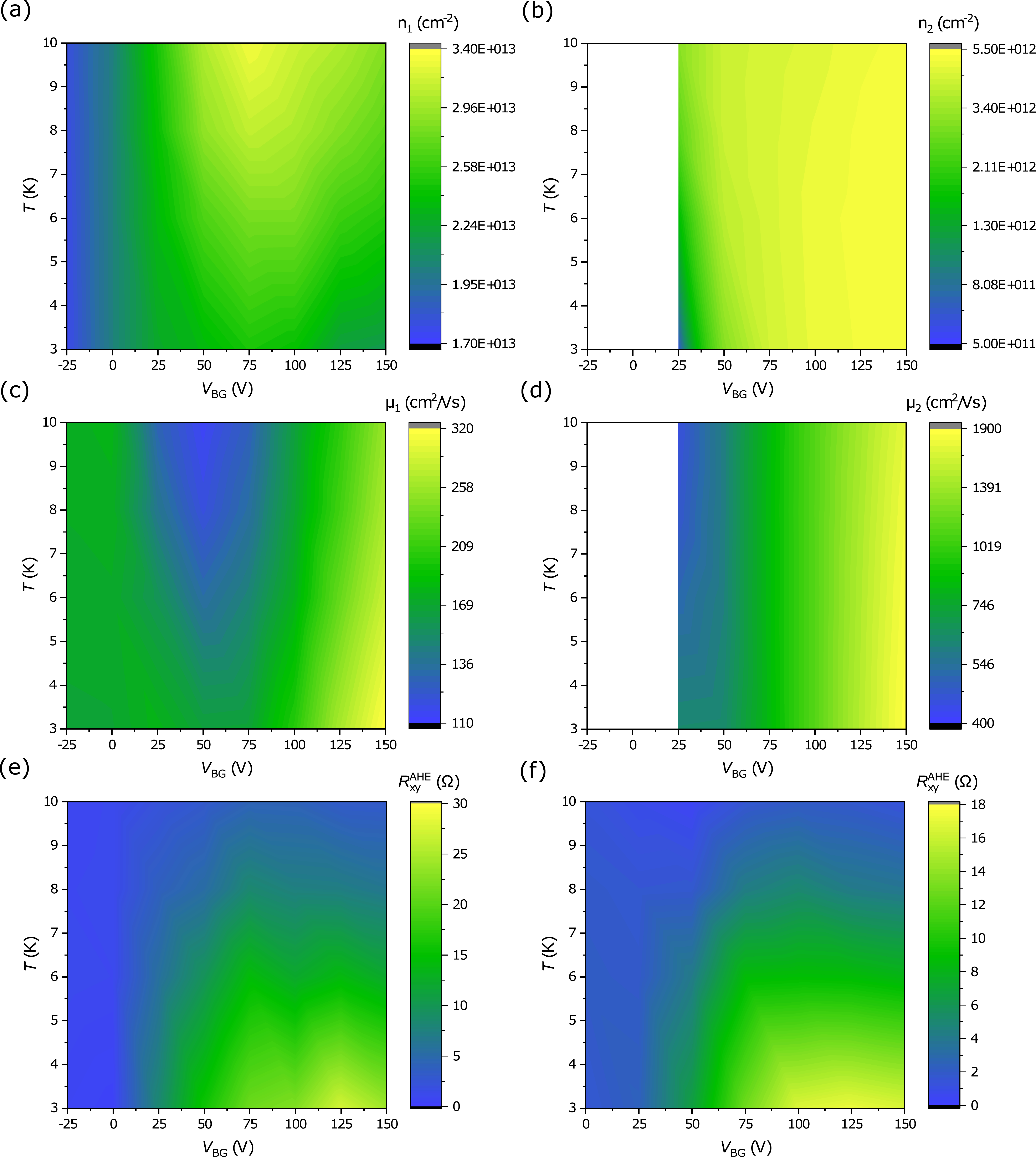}
\caption{Temperature and back gate voltage dependence of (a-b) carrier concentrations, (c-d) mobilities, (e) Anomalous Hall coefficient for the LAO/ETO/STO sample. (f) Anomalous Hall coefficient for the LAO/ELTO/STO sample. The color scale for all quantities gives the range of their values.}\label{Fig5-ETO_Vgate}
\end{figure*}

\subsection{Gate tuning of the magnetotransport properties }

Next, for all samples, we apply a series of gate voltages V$_G$, starting at +150~V and going down to -150~V in a number of steps, and vary the temperature. We observe the development of the minimum in R$_S$(T) below +25~V in magnetically doped samples (Fig.~\ref{Fig4-ETO_gating}a,b), which becomes more pronounced with increasingly negative V$_G$. That particular feature in the temperature dependence is often attributed to Kondo-like behaviour\rec{NMATLAOETO}, which might well be connected to the magnetism of the doping layers. However, and surprisingly, we see quite similar behavior in nominally non-doped LAO/STO samples (Fig.~\ref{Fig4-ETO_gating}c), where we do not observe the AHE. We also note, in all cases, that $R_S(T)$ tends to saturate at low temperatures, which is not in line with the expected logarithmic increase of the Kondo resistance. These are intriguing observations and the reason to look more closely at the actual effects of the magnetic interlayers.\\
For this we studied the combined temperature dependence and gate dependence of the AHE in doped samples below 10~K, which also entails determining $n_i, \mu_i$. In order to represent both temperature- and gate voltage-dependence, we use a false-color plot, where the colors represent the values of the various paramaters. This is done in Fig.~\ref{Fig5-ETO_Vgate}(a-e) for LAO/ETO/STO. All in all, the results are similar to those of previous AHE studies in the two-band regime in the oxide interfaces\rec{JoshuaNCOM,NMATLAOETO,PRBWLETO,GanPRB}. We find a Lifshitz point\rec{JoshuaNCOM} located near 25 V as can be seen from the behavior of $n_2$ (Fig.~\ref{Fig5-ETO_Vgate}b, where carriers of the second type disappear below that voltage. Unlike the carrier concentration and mobility of the low mobile carriers (Fig.~\ref{Fig5-ETO_Vgate}a,c), the carrier concentration and mobility of the high mobile carriers (Fig.~\ref{Fig5-ETO_Vgate}b,d) are less sensitive to a change in temperature. On the other hand, the change in voltage affects $n_i$ and $\mu_i$ much stronger. The behavior is qualitatively the same for both samples; numbers for LAO/ELTO/STO can be found in the Supplement Fig.~S4. \\
The AHE (Fig.~\ref{Fig5-ETO_Vgate}e) is not present in the region where two-band behavior is absent, as would be expected in the picture where the AHE is simply controlled by the second type of carriers\rec{JoshuaPNAS}. However, \Rahe{} also disappears above 10~K. That can be due to vanishing of the magnetization\rec{NMATLAOETO}, of the spin-orbit coupling\rec{GTOpaper}, or both. Indeed we see the signature of weak antilocalization (WAL), indicative of spin orbit coupling in the low field MR in all samples (Supplement Fig.~S5). Note that the MR is strongly affected by changes in gate voltage and temperature (see Supplement Fig.~S5). At the same time, the AHE coefficient is higher in the ETO-based sample than in the ELTO-based sample. Apart from that, the gate voltage and temperature dependence of the AHE coefficient for LAO/ELTO/STO (Fig.~\ref{Fig5-ETO_Vgate}f) are similar to LAO/ETO/STO. Some notes on the mechanism of the AHE can be found in supplementary. Here, we point out that the upturn in R$_S$(T) happens at the same voltage where \Rahe{} and $n_2$ fall to zero in the magnetically doped samples. So, magnetic doping significantly affects the spin-polarized phase of the 2DEG in the two-band regime, where the moment was argued to be ferromagnetically coupled with \dxyz{} electrons, but seems not to have any qualitative effect on the Kondo-like (one-band) regime, where moments has been claimed to be antiferromagnetically coupled with \dxy{} electrons\rec{JoshuaPNAS, RuhmanPRB}.

\begin{figure*}
\includegraphics[scale=0.35]{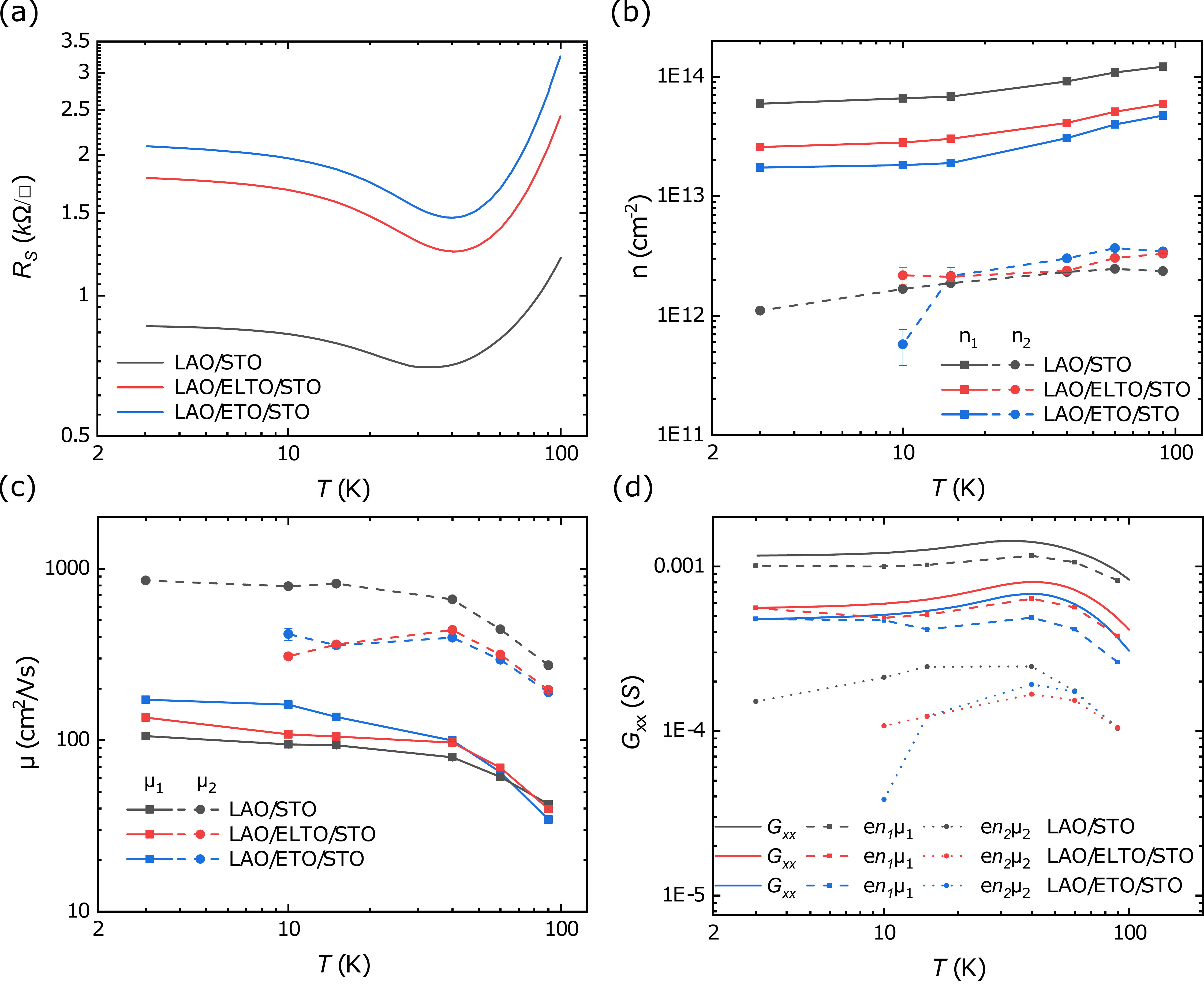}
\caption{Measurements at a gate voltage of -100~V for all three samples. (a) the temperature-dependent sheet resistance $R_S(T)$, (b) the carrier concentrations; note the disappearance of the second band below 10~K for the E(L)TO samples (c) the mobilities and (d) the conductance.}\label{Fig6-ETO_100V}
\end{figure*}

To understand the nature of this behavior, we measured the temperature dependence of the sheet resistance $R_S(T)$ for all three samples at -100~V in a larger temperature range (from 100~K downward) and performed magnetotransport measurements during cooldown at a few temperatures in order to find carrier concentrations and mobilities at this gate voltage. As can be seen in Fig.~\ref{Fig6-ETO_100V}a, the samples show different values of the sheet resistance but follow the same trend with a minimum around 30-40 K and saturation-like behavior below 10~K. The extracted carrier concentrations (Fig.~\ref{Fig6-ETO_100V}b) from the magnetotransport data showed a constant decrease in carrier density with decreasing temperature. This is similar to the behavior of the carrier concentration of the low mobility carriers without back gate voltage, but here it is also observed for the high mobility carriers, and especially strong below 10 K. At the same time, the mobility (Fig.~\ref{Fig6-ETO_100V}c) increases at first with decreasing temperature, but below 40 K saturates, similar to the temperature dependence without applied back gate voltage, and can be described the same scattering picture. Important to note is that the change in the temperature dependence of the mobility happens around the temperature corresponding to the minimum in the sheet resistance. Finally, to illustrate how the change in carrier concentrations and mobilities affects the temperature dependence of the sheet resistance, we plot conductance $G_{xx}=1/R_{S}$ and band conductances $en_i\mu_i$ in Fig.~\ref{Fig6-ETO_100V}

\section{Discussion\label{Discussion}}
The results shown here are consistent with studies of LAO/ETO/STO in Ref. \onlinecite{LucaPRB,NMATLAOETO,PRBWLETO}. The same as in the earlier studies, we observe the MIT, the tunable AHE, the appearance of a resistance minimum and also WAL behavior. The AHE is the possible signature of ferromagnetism. Indeed, XMCD data seems to support that picture\rec{LucaPRB,NMATLAOETO}. Nevertheless, bulk ETO is an antiferomagnet\rec{ChienPRB,McGuireJAP,SchiemerPRB,BHPRB,BessasPRB,EllisPRB,KatsufujiPRB}. It is also isostructural to STO, and, therefore, stoichiometric ETO should remain antiferromagnetic, when grown on STO\rec{LeeJHAPL}. At the same time, experiments revealed that antiferromagnetism only occurs for post-annealed PLD films\rec{ShimamotoAPL,FujitaAPL}, and films grown by MBE\rec{LeeJHAPL}. Here we used PLD films without annealing, which can become ferromagnetic\rec{FujitaAPL,ShimamotoAPL, TanakaJMR,KUGIMIYAJMMM} because of either the formation of oxygen vacancies\rec{ShimamotoAPL, ELTOSCAD} or a longer out-of-plane lattice constant, which leads to a bigger lattice volume\rec{FujitaAPL, TanakaJMR, YLinACSAPMI}. The films tend to be ferromagnetic if the ratio between out-of-plane constant and in-plane constant is larger than 1.02 or less than 0.99. Due to large error bars, we cannot estimate this ratio precisely, but our films may be on the edge of a ratio of 1.02~\rec{YLinACSAPMI}. The crystallinity of our films allows us to exclude amorphisation of ETO as the driving mechanism for the transition to a ferromagnetic state\rec{AkamatsuPRB}. The doping with La\rec{KatsufujiPRBELTO,ELTOPRL, ELTOSCAD, ShinPRB} can also turn ETO into a ferromagnetic metal. The observed MIT and its weak sensitivity to the La content indicate that the occurrence of the AHE has a more complicated origin than just due to ferromagnetism induced by off-stochiometry in the ETO layer. However, that scenario it hard to exclude due to intermixing in samples. Because it seems that we overshoot the sweet spot of 10~~\% for La doping, extra doping may weaken ferromagnetism\rec{ShinPRB}. \\
Ref.\onlinecite{GunkelPRX} observed variations of the AHE with changes in the oxygen pressure used during the growth and proposed an indirect connection between the AHE and magnetism induced by Sr vacancies rather than with actual magnetic moments, including \tit{} induced by oxygen vacancies. However, we used the same growth pressure for the non-doped LAO/STO sample and LAO/E(L)TO/STO, so the variation of Sr vacancies should not be too high from sample to sample. With introducing Eu doping, we observe an increase of the AHE magnitude contrary to Ref.\onlinecite{GunkelPRX}.\\ Another defect scenario concerns the formation of B-site cation defects in the LAO layer\cite{ParkNCOM}. Indeed we observe in our samples the presence of Ti deeper in the LAO layer, which can give an additional contribution to the magnetism. Refs.\onlinecite{RuhmanPRB, JoshuaPNAS} proposed a ferromagnetic coupling of \dxyz{} with localized magnetic moments such as \tit{} formed due to oxygen vacancies. That indeed can explain the observation of AHE above Lifshitz point, since skew-scattering of \dxyz{} carriers seems to be a dominant contribution to the AHE in oxide interfaces in the two-band regime\rec{JoshuaPNAS, GunkelPRX, NMATLAOETO, ZhangLSMOPRB}. Nevertheless, it appears to be not straightforward to explain our results on the tunable AHE. First, the AHE coefficient does not scale linearly with the mobility of high mobile carriers as shown in supplementary. Second, and probably more importantly, our non-doped LAO/STO sample does not exhibit an AHE, despite the presence of \dxyz{} carriers. All in all, there are certainly multiple ways to induce magnetization and AHE, but in our samples magnetic doping is the main one.

In the framework of magnetic doping, we have further indication that the picture of magnetic interactions based purely on the symmetry of orbitals does not hold, specifically from the data in the negative gate voltage range, where the AHE vanishes. The authors of Ref. \onlinecite{RuhmanPRB, JoshuaPNAS} argued that the coupling between mobile \dxy{} electrons and localized magnetic moments is antiferromagnetic. Strong support of this picture came from the behavior of the negative in-plane MR. Above the Lifshitz point, the observed large drop of (negative) in-plane MR was attributed to the destruction of Kondo-screening and the polarisation of magnetic moments with applied field\rec{BenShalomPRB, RuhmanPRB, JoshuaPNAS}. However, that picture was challenged by Diez {\it et al.}\rec{DiezPRL} who found that the negative in-plane MR survives up to $20$ K, opposite to the expectations from the temperature-dependent Kondo picture. Furthermore, single-particle Boltzmann transport theory was sufficient to reproduce the large negative MR by taking into the account finite-range electrostatic impurity scattering, the anisotropic Fermi surface above the Lifshitz transition and the SOC. \\

In this work, we also challenge the Kondo picture. The experiments on the LAO/STO interface performed in Ref. \onlinecite{FuchsPRB} showed that the resistance minimum could be controlled by applying hydrostatic pressure. They concluded that impurity scattering, the pressure, and the temperature dependence of the STO dielectric constant plus thermally activated charge trapping form the mechanism responsible for the resistance minimum. Although we studied the back gate voltage dependence of the sheet resistance in this work, our result can be interpreted in a somewhat similar manner. At negative gate voltages, the carrier concentration decreases in the whole temperature range, but down to 40-50 K the mobility enhances due to a decrease of electron-electron scattering. Taken together, that is the reason why we observe a decrease in the resistance down to a minimum temperature. However, with a further decrease in temperature, the change in mobility is much less steep due to various scattering mechanisms coming into play, as discussed in Sec.~\ref{ETO_MT}.
Simultaneously, the decrease of the carrier concentration is continuous and especially pronounced for the high mobility carriers. It is not surprising then that in this region the sheet resistance starts to grow again.

Two mechanisms are responsible for the decrease in the carrier concentration. One is charge trapping\rec{BiscarasSREP,ChunhaiPRL}, which is less effective at high temperatures than at low temperatures. Second and probably dominant here is the complex behavior of the electric permittivity of the STO single crystal substrate at low temperatures. Indeed, at high temperatures above 40 K, the permittivity hardly changes with the applied electric field, but it does change significantly at lower temperature\rec{BethePRR, VendikJOS, NevilleJAP}. It means that the effect of back gating will be significantly more efficient in the low-temperature region.
Furthermore, it is well known that back gate experiments significantly affect the carrier concentration of the second type of carriers\rec{JoshuaNCOM, BiscarasPRL} at low temperatures. Below 10 K, a back gate voltage depletes, or almost depletes, the second band; therefore the saturation-like behavior of $R_S$ in this range can be explained by the much weaker change in the mobility and carrier concentration of the low mobile carriers with temperature. Note, that our results do not exclude a possibility to induce the Kondo effect at the LAO/STO interface, but only indicate that in the back gate geometry, other effects are dominant and responsible for the resistance minimum in the temperature dependence when it occurs after the application of a gate voltage. Other experiments, and specifically top gating,  may be the better way to study the Kondo effect in the oxide heterostructures.\rec{YanPRB}.

\section{Conclusion\label{Conclusion}}
In conclusion, we have studied the LAO/STO interface with and without an inserted magnetic delta-doping layer of \ELTO{} (either through La intermixing or intentionally). Our experimental results are mostly in agreement with previous research on these structures\rec{LucaPRB,NMATLAOETO, PRBWLETO}. We observed that, despite extra Lanthanum, the MIT happens at 2 u.c. Moreover, the AHE is indeed induced by inserting the 2 u.c. of ETO or ELTO, and can be gate tuned. Furthermore,  at negative back gate voltages we find minima in the sheet resistance as function of temperature. However, we do not attribute this to Kondo-like behavior, the presence of magnetic moments notwithstanding. In our interpretation, the minima rather arise from the temperatuere dependence of the carrier concentrations and mobilities of both conduction bands, and is most probably due to spinless scattering mechanisms, charge trapping, and the complex temperature and electric field dependence of the STO permittivity.

\begin{acknowledgments}
N.L. and J.A. gratefully acknowledge the financial support of the research program DESCO, which is financed by the Netherlands Organisation for Scientific Research (NWO). J.V. and N.G. acknowledge funding from the Geconcentreerde Onderzoekacties (GOA) project “Solarpaint” of the University of Antwerp and the European Union’s horizon 2020 research and innovation programme under grant agreement \textnumero 823717 - ESTEEM3. The Qu-Ant-EM microscope used in this study was partly funded by the Hercules fund from the Flemish Government. The authors want to thank M. Stehno, G. Koster, and F.J.G. Roesthuis for useful discussions.
\end{acknowledgments}

\bibliography{ETO.bib}
\end{document}